\documentclass[10pt]{article}
\usepackage{amssymb}
\usepackage{amsmath}
\usepackage{amsthm}
\usepackage{latexsym}
\usepackage[dvips]{epsfig}
\usepackage{mathrsfs}
\usepackage[bookmarks,colorlinks=false]{hyperref}

\theoremstyle{plain}
\newtheorem{proposition}{Proposition}
\newtheorem{remark}{Remark}

\newtheorem{theorem}{Theorem}

\newcounter{mnotecount}[section]

 \newcommand{\mnotex}[1]
 {\protect{\stepcounter{mnotecount}}$^{\mbox{\footnotesize $\bullet$\themnotecount}}$ 
 \marginpar{
 \raggedright\tiny\em
 $\!\!\!\!\!\!\,\bullet$\themnotecount: #1} }

\begin{document}

\title{\textbf{New conserved currents for vacuum space-times in dimension four 
with a Killing vector}}

\author{{\Large Alfonso Garc\'{\i}a-Parrado G\'omez-Lobo} \thanks{E-mail address:
{\tt alfonso@math.uminho.pt}} \\
F\'{\i}sica Te\'orica, Universidad del Pa\'{\i}s Vasco,
Apartado 644, 48080 Bilbao, Spain,\\
Centro de Matem\'atica, Universidade do Minho,
4710-057 Braga, Portugal.
}
\maketitle

\begin{abstract}
A new family of conserved currents for vacuum space-times with a Killing vector is presented. The currents are 
constructed from the superenergy tensor of the Mars-Simon tensor and using the positivity properties of the
former we find that the conserved charges associated to the currents have natural positivity properties in certain cases. Given 
the role played by the Mars-Simon tensor in local and semi-local characterisations of the Kerr solution,
the currents presented in this work are useful to construct non-negative scalar quantities characterising 
Kerr initial data (known in the literature as {\em non-Kerrness}) which in addition are {\em conserved charges}.
\end{abstract}

PACS: 04.20.Jb, 11.40.-q, 04.20.-q\\

MSC: 83C15, 83C05, 58A25

\section{Introduction}
\label{sec:intro}
The knowledge about conserved currents in mathematics and physics has paramount relevance in different contexts
such as proofs of uniqueness for partial differential equations
and the construction of global conserved quantities. In general relativity 
a natural situation where this arises is when the space-time 
admits a Killing vector field.
Perhaps the most known example is the contraction of the Killing vector with 
the energy-momentum tensor. In this spirit, an alternative possibility arises by using  
the so-called super-energy tensors because they have natural positivity properties which 
can be translated to the corresponding conserved charges (see  
\cite{LAZKOZ-VERA-SENO,ERIK-BERG-SENO,ERIKSSON-CURRENT-I,ERIKSSON-CURRENT-II} for explicit examples and 
applications). 
In this paper we follow this approach and
present a new family of conserved currents for a vacuum space-time with a Killing vector 
based on the super-energy of the {\em Mars-Simon tensor}. We show that 
the family can be parametrised in terms of a complex constant.
The Mars-Simon tensor 
has the same algebraic properties as the Weyl tensor and hence its superenergy tensor is similar to the Bel-Robinson
tensor which is the superenergy tensor of the Weyl tensor (see \cite{SUPERENERGY} and references therein).
By using the Bel-Robinson tensor one can construct for any vacuum
space-time admitting a Killing vector the {\em Bel-Robinson current} and we show by means of an explicit example 
that our current and Bel-Robinson's are in general independent. 
Therefore our current has an independent interest to the Bel-Robinson one. 

A consequence of the positivity properties of the superenergy tensor is that the conserved charge associated to our 
current is automatically positive if the Killing vector is time-like (stationary space-time) and the hypersurface space-like
or null. We exploit these properties to construct a non-negative quantity $\mathcal{Q}(\Sigma)$ on a space-like
embedded hypersurface $\Sigma$ which, under certain conditions vanishes if and only if $\Sigma$ is a Cauchy hypersurface of an open 
subset in a stationary region of the Kerr solution and it has 
the additional property that it is a conserved charge because it arises from a conserved current 
(see Theorem \ref{theo:conserved-charge} for full details).

The paper is structured as follows: in section \ref{sec:mars-simon} we recall the main properties of the Mars-Simon tensor.
Section \ref{sec:conserved-currents} presents the main result of the paper, namely, the construction of the conserved
current using the superenergy of the Mars-Simon tensor (Theorem \ref{theo:ms-conserved-current}). We use spinors in the proof of the conservation so the elements 
of spinor calculus which we need are also reviewed. Finally in section \ref{sec:applications} we point out some possible 
applications to the characterisation of Kerr initial data through non-negative quantities ({\em non-Kerrness}). 
The algebraic computations of this paper have been carried out with the suite {\em xAct} \cite{XACT}.

\section{The Mars-Simon tensor in vacuum}
\label{sec:mars-simon}
We shall work in a four dimensional smooth Lorentzian manifold $(\mathcal{M}, g_{\mu\nu})$ 
and use small Greek letters to denote abstract tensor indices. Occasionally we shall employ index free
notation in which case we will use boldface kernel letters for tensors.
Round (square) indices enclosing a set of abstract indices denote symmetrization (anti-symmetrization).
Our convention for the metric signature is 
$(+,-,-,-)$ and the notation for the Riemann and Ricci tensors defined from the 
Levi-Civita connection $\nabla_\mu$ is the standard one, 
$R^\mu{}_{\nu\alpha\beta}$, $R_{\mu\nu}\equiv R^\mu{}_{\nu\mu\beta}$. In this paper we assume that the 
Lorentzian manifold $\mathcal{M}$ (space-time) is a vacuum solution of 
the Einstein field equations, $R_{\mu\nu}=0$ and therefore the Riemann tensor is the same as the 
Weyl tensor $W_{\alpha\beta\mu\nu}$. When working with complex quantities we use an overbar for the complex 
conjugate.

We recall some well-known formulae for a vacuum space-time $\mathcal{M}$ admitting a Killing vector. 
Proofs are omitted and the reader is referred to the literature for further 
details \cite{HEUSLER,MARS-EHLERS}. A Killing vector field $\vec{\boldsymbol \xi}$ (Killing 1-form)
is defined by the differential condition 
\begin{equation}
\pounds_{\vec{\boldsymbol \xi}}g_{\mu\nu}=0\;,\quad
\nabla_\mu \xi_\nu+\nabla_\nu\xi_\mu=0.
\label{eq:Killing-condition}
\end{equation}
The Killing condition enables us to define the {\em Killing 2-form}
\begin{equation}
F_{\mu\nu}\equiv \nabla_\mu\xi_\nu\;,
\label{eq:killin-2-form}
\end{equation}
which satisfies the following differential identity in vacuum
\begin{equation}
\nabla_\beta F_{\alpha\chi}=-W_{\alpha\chi\beta\rho}\xi^\rho.
\label{eq:covdiv-F}
\end{equation}
Many equations in this work are better expressed in the {\em complex formalism}. In 
this formalism one defines for the Weyl tensor and Killing 2-form their {\em self-dual} 
counterparts
\begin{equation}
 \mathcal{F}_{\alpha\beta}\equiv F_{\alpha\beta}+\rm{i}\; F^*_{\alpha\beta}\;,\quad
 \mathcal{W}_{\alpha\beta\chi\rho}\equiv W_{\alpha\beta\chi\rho}+\rm{i}\; W^*_{\alpha\beta\chi\rho}\;, 
\label{eq:self-dual-tensors}
\end{equation}
where the Hodge duality operation is given by
\begin{equation}
 F^*_{\alpha\beta}\equiv \frac{1}{2}\eta_{\alpha\beta}{}^{\chi\rho}F_{\chi\rho}\;,\quad
 W^*_{\alpha\beta\chi\rho}\equiv \frac{1}{2}\eta_{\chi\rho}{}^{\mu\nu}W_{\alpha\beta\mu\nu}\;,\quad
 {}^*W_{\alpha\beta\chi\rho}\equiv \frac{1}{2}\eta_{\alpha\beta}{}^{\mu\nu}W_{\mu\nu\chi\rho}\;,\quad
\label{eq:dual-tensors}
\end{equation}
and $\eta_{\mu\nu\alpha\beta}$ is the volume form. One has the algebraic properties
\begin{equation}
\mathcal{F}^*_{\mu\nu}=-\rm{i}\; \mathcal{F}_{\mu\nu}\;,\quad
\mathcal{W}^*_{\alpha\beta\mu\nu}=-\rm{i}\; \mathcal{W}_{\alpha\beta\mu\nu}=
{}^*\mathcal{W}_{\alpha\beta\mu\nu}\;,
\end{equation}
and the differential identity
\begin{equation}
\nabla_\beta \mathcal{F}_{\alpha\chi}=-\mathcal{W}_{\alpha\chi\beta\rho}\xi^\rho.
\label{eq:cd-selfdualF}
\end{equation}
The Ernst 1-form is defined by
\begin{equation}
\sigma_\alpha\equiv 2 \mathcal{F}_{\alpha\beta}\xi^\beta\;, 
\label{eq:ernst-form}
\end{equation}
and in a vacuum space-time it is closed enabling us to define a local potential $\sigma$ 
(the Ernst potential)
\begin{equation}
\nabla_{[\mu}\sigma_{\nu]}=0\;,\quad \sigma_\mu=\nabla_\mu\sigma.
\end{equation}
The Ernst potential can be written in terms of the {\em Killing norm} $\lambda$ and {\em twist} 
$\omega$ as follows
\begin{equation}
\sigma=k+\lambda+2\ \rm{i}\ \omega\;,\quad k\in\mathbb{C}\;,\quad
\lambda\equiv \xi_\mu\xi^\mu.
\label{eq:define-ernst-potential}
\end{equation}
 
With the above definitions it is possible to introduce a rank-4 tensor at any point 
where the Ernst potential does not vanish. Its explicit definition is \cite{MARS-KERR,IONESCU-KLAINERMAN}
\begin{equation}
\mathcal{S}_{\alpha\beta\chi\mu}\equiv \mathcal{W}_{\alpha\beta\chi\mu}+
\frac{6}{\sigma}\left(\mathcal{F}_{\alpha\beta}\mathcal{F}_{\chi\mu}
-\frac{\mathcal{F}_{\kappa\rho}\mathcal{F}^{\kappa\rho}}{3}\mathcal{I}_{\alpha\beta\chi\mu}\right)\;,
\label{eq:define-ms}
\end{equation}
where $\mathcal{I}_{\alpha\rho\beta\chi}$ is the so-called metric in the space of the self-dual 2-forms
\begin{equation}
{\mathcal I}_{\alpha\rho\beta\chi}\equiv
\frac{1}{4}(\rm{i}\; \eta_{\alpha\rho\beta\chi}+g_{\alpha\beta}g_{\rho\chi}-
g_{\alpha\chi}g_{\beta\rho}).
\label{eq:2-form-metric}
\end{equation}
The tensor $\mathcal{S}_{\alpha\beta\chi\mu}$ is called the {\em Mars-Simon tensor} and it has been 
studied by a number of authors 
\cite{IONESCU-KLAINERMAN,GARSENKERRNESS}. Its main relevance is that it can be used 
in the formulation of local and semi-local 
characterizations
of the Kerr solution \cite{MARS-KERR,MARS-KERR-UNIQUENESS}. For us the most important properties of the the Mars-Simon tensor are that 
it is a {\em Weyl candidate}
\begin{equation}
\mathcal{S}_{[\mu\nu]\alpha\beta}=\mathcal{S}_{\mu\nu\alpha\beta}\;,\quad
\mathcal{S}_{\mu\nu\alpha\beta}=\mathcal{S}_{\alpha\beta\mu\nu}\;,\quad
\mathcal{S}_{[\mu\nu\alpha]\beta}=0\;,
\end{equation}
it is self dual 
\begin{equation}
\rm{i}\; \mathcal{S}^*_{\mu\nu\alpha\beta}=\mathcal{S}_{\mu\nu\alpha\beta}\;,
\label{eq:ms-self-dual}
\end{equation}
and it also fulfills the differential identity 
\begin{equation}
\sigma\nabla_\alpha\mathcal{S}^\alpha_{\phantom{\alpha}\beta\gamma\delta}
=2\xi^\alpha(2(\mathcal{F}^\rho_{\phantom{\rho}[\delta}\mathcal{S}_{\gamma]\beta\alpha\rho}+
\mathcal{F}_\beta^{\phantom{\beta}\rho}\mathcal{S}_{\delta\gamma\alpha\rho})
+g_{\beta[\delta}\mathcal{S}_{\gamma]\alpha\rho\pi}\mathcal{F}^{\rho\pi}). 
\label{eq:covdiv-ms}
\end{equation}
See \cite{IONESCU-KLAINERMAN,GAR-VAL-KERR,GARSENKERRNESS} for an account of all these properties of the Mars-Simon tensor.

\section{A new family of conserved currents in vacuum}
\label{sec:conserved-currents}
We show in this section how to construct a set of conserved currents for a vacuum space-time 
out of the Mars-Simon tensor.
In the computations which follow we will make extensive use of the spinor formalism 
and hence we assume that the manifold $\mathcal{M}$ admits 
a {\em spin structure}. 
Since these computations are {\em local} this is not really a restriction.
In our set-up the spin structure is represented by a smooth field 
$\sigma^\mu{}_{AA'}$ (soldering form) which serves to relate spinors and tensors 
by transforming pairs of spinor indices into 
tensor ones and back. The field $\sigma^\mu{}_{AA'}$ is globally defined on $\mathcal{M}$ if and only if 
$\mathcal{M}$ admits a spin structure \cite{ASHTEKAR} but as said we do not need this requirement
to establish our results.
One constructs the spin bundle in the standard way and we 
follow the traditional convention of using capital Latin letters for abstract spinor indices  
(which can be unprimed or primed). Our conventions for 
the definition of the spin metric $\epsilon_{AB}$, spin connection $\nabla_{AA'}$ and curvature spinors
follow those of \cite{PR-RINDLER-1}. When relating spinors and tensors we will omit the soldering form in the 
formulae unless a confusion might arise. 

Since we work in vacuum, the only non-vanishing part of the curvature is the one corresponding to the Weyl tensor
which as is well-known can be represented by a totally symmetric spinor $\Psi_{ABCD}$ (Weyl spinor) fulfilling the 
spinor form of the Bianchi identity
\begin{equation}
\nabla^{A}{}_{A'}\Psi_{ABCD}=0. 
\label{eq:spinor-bianchi}
\end{equation}
For later use we also recall the spinor form of the volume element
\begin{equation}
\eta_{AA'BB'CC'DD'}=
i (\epsilon_{AC} \epsilon_{BD} \epsilon_{A'D'} \
\epsilon_{B'C'} -  \epsilon_{AD} \epsilon_{BC} \
\epsilon_{A'C'} \epsilon_{B'D'}).
\label{eq:spinor-volumeform}
\end{equation}

To proceed we need to express the tensor formulae shown in section \ref{sec:mars-simon} in the spinor language.
The Killing vector $\vec{\boldsymbol{\xi}}$ has the spinor form $\xi^{AA'}$ and the Killing 2-form $F_{\mu\nu}$
admits the spinor decomposition
\begin{equation}
F_{AA'BB'}\equiv\nabla_{AA'}\xi_{BB'}=\epsilon_{AB}\bar{\phi}_{A'B'} + \epsilon_{A'B'}\phi_{AB}\;,
\label{eq:killing-2-form-spinor}
\end{equation}
where $\phi_{AB}$ is a totally symmetric spinor given by
\begin{equation}
\phi_{AB}\equiv\frac{1}{2}\nabla_{(A|B'|}\xi_{B)}{}^{B'}=\frac{1}{2}\nabla_{AB'}\xi_{B}{}^{B'}\;, 
\label{eq:killing-2-form-spinor-symmetric}
\end{equation}
and in the last step we used the spinor form of the Killing condition (\ref{eq:Killing-condition})
\begin{equation}
 \nabla_{AA'}\xi_{BB'}+\nabla_{BB'}\xi_{AA'}=0.\label{eq:spinor-killing-condition}
\end{equation}

Once the spinor expression of $F_{\mu\nu}$ is known we can easily compute  
the spinor expression of $\mathcal{F}_{\mu\nu}$ if we use (\ref{eq:spinor-volumeform}). The result is
\begin{equation}
\mathcal{F}_{AA'BB'}=2\epsilon_{A'B'}\phi_{AB}.
\label{eq:spinor-selfdualF}
\end{equation}
We can now write (\ref{eq:covdiv-F}) (or equivalently (\ref{eq:cd-selfdualF})) in terms of spinors as
\begin{equation}
\nabla_{CC'}\phi_{AB}=\Psi_{ABCD}\xi^D{}_{C'}. 
\label{eq:spinor-cdphi}
\end{equation}
Combining (\ref{eq:spinor-selfdualF}) with (\ref{eq:ernst-form}) we obtain the spinor expression of 
the Ernst 1-form
\begin{equation}
 \sigma_{AA'}=-4\xi^B{}_{A'}\phi_{AB}.
\label{eq:spinor-ernst-1-form}
\end{equation}
The quantity $\mathcal{I}_{\alpha\rho\beta\chi}$ can be also easily written in spinor form if 
we use (\ref{eq:spinor-volumeform}). The result is
\begin{equation}
\mathcal{I}_{AA'CC'BB'DD'}=\frac{1}{4}(\epsilon_{AD}\epsilon_{BC}-\epsilon_{AB}\epsilon_{CD})
\epsilon_{A'C'}\epsilon_{D'B'}.
\label{eq:I-spinor}
\end{equation}

We have now all the ingredients necessary to obtain the spinor expression of the Mars-Simon tensor. 
One takes
the spinor forms of $\mathcal{W}_{\mu\nu\alpha\beta}$, $\mathcal{F}_{\mu\nu}$, and 
$\mathcal{I}_{\alpha\beta\mu\nu}$ found above and replaces them into the definition of the Mars-Simon tensor 
(\ref{eq:define-ms}). After some algebra one gets
\begin{equation}
\mathcal{S}_{AA'BB'CC'DD'}=2\epsilon_{A'B'}\epsilon_{C'D'}\left(\Psi_{ABCD}+\frac{2}{\sigma}
\left(6\phi_{AB}\phi_{CD}-(\epsilon_{AD} \epsilon_{BC} + \epsilon_{AC} \epsilon_{BD})\phi_{FH}\phi^{FH}\right)
\right). 
\end{equation}
This expression can be further simplified if we use the identity
\begin{equation}
\phi_{AB} \phi_{CD} = \phi_{(AB} \phi_{CD)} + \frac{\phi_{FH} \phi^{FH}}{6} (\epsilon_{AD} \
\epsilon_{BC} + \epsilon_{AC} \epsilon_{BD}).
\end{equation}
The final result is
\begin{equation}
\mathcal{S}_{AA'BB'CC'DD'}=2\epsilon_{A'B'}\epsilon_{C'D'}\left(\Psi_{ABCD}+\frac{12}{\sigma}
\phi_{(AB} \phi_{CD)}\right).
\end{equation}
Bearing in mind the previous relation we define the Mars-Simon spinor by
\begin{equation}
\mathcal{S}_{ABCD}\equiv \Psi_{ABCD}+\frac{12}{\sigma}\phi_{(AB} \phi_{CD)}.
\label{eq:mars-simon-spinor}
\end{equation}
The Mars-Simon spinor is totally symmetric as is evident from its definition. Another important
property of the Mars-Simon spinor is given next.
\begin{proposition}
The Mars-Simon spinor fulfills the following differential identity
\begin{equation}
\nabla_{AA'}\mathcal{S}_{BCD}{}^A=\frac{12}{\sigma}\xi^{A}{}_{A'}\mathcal{S}_{FA(CD}\phi_{B)}{}^F.
\label{eq:covdiv-ms-spinor}
\end{equation}
\label{prop:spinor-marssimon-covdiv}
\end{proposition}

\proof 
This is a computation involving (\ref{eq:mars-simon-spinor}), (\ref{eq:spinor-cdphi}), 
the Bianchi identity (\ref{eq:spinor-bianchi}), the definition of the Ernst potential written in
spinor form
\begin{equation}
\nabla_{AA'}\sigma=\sigma_{AA'}\;,
\label{eq:spinor-ernst-potential}
\end{equation}
and (\ref{eq:spinor-ernst-1-form}).
The result of the computation is
\begin{equation}
\nabla_{AA'}\mathcal{S}_{BCD}{}^{A} = \frac{12 \xi^{A}{}_{A'}}{\sigma^2}
\bigl(-4 \phi_{(BC}\phi_{DF)} \phi_{A}{}^{F} + \sigma  \Psi_{AF(CD} \phi_{B)}{}^{F}
\bigr)\;,
\label{eq:covdiv-ms-intermediate}
\end{equation}
where we used the identity
\begin{equation}
\nabla^{A}{}_{A'}\phi_{(AB} \phi_{CD)}=
-\Psi_{AF(CD}\phi_{B)}{}^{F}\xi^{A}{}_{A'}.
\end{equation}
To complete the proof we express all the occurrences of the Weyl spinor in
(\ref{eq:covdiv-ms-intermediate}) in terms of the Mars-Simon spinor using 
(\ref{eq:mars-simon-spinor}).\qed
\subsection{The superenergy of the Mars-Simon spinor}
The Mars-Simon spinor is totally symmetric and hence we can easily compute its {\em superenergy}
\begin{equation}
 T_{AA'BB'CC'DD'}\equiv 4\ \mathcal{S}_{ABCD}\bar{\mathcal{S}}_{A'B'C'D'}. 
\label{eq:ms-superenergy}
\end{equation}
The spinor $T_{AA'BB'CC'DD'}$ is Hermitian and it corresponds to the superenergy tensor constructed
from the Mars-Simon tensor $\mathcal{S}_{\mu\nu\rho\sigma}$ \cite{BERGQVIST-SPINOR-SUPERENERGY}. Its tensor counterpart, 
$T_{\alpha\beta\gamma\delta}$, is given by
\begin{equation}
 T_{\alpha\beta\gamma\delta}\equiv \mathcal{S}_{\alpha\phantom{\mu}\phantom{\nu}\beta}^{\phantom{\alpha}\mu\nu}
 \bar{\mathcal{S}}_{\gamma\mu\nu\delta}\;,
\end{equation}
and it
was used in \cite{GARSENKERRNESS} to define {\em quality factors} which measures the {\em proximity} of a 
solution to the Kerr solution. In this work we show another application
of this superenergy tensor consisting in the construction of a family of conserved currents.
\begin{theorem}
For any vacuum solution of the Einstein's field equations $(\mathcal{M}, g_{\mu\nu})$ admitting a 
Killing vector field consider an open 
subset $\mathcal{U}\subset\mathcal{M}$ where the Ernst potential is differentiable and  
non-vanishing and 
define the tensor $T_{\alpha\beta\gamma\delta}$
as explained above. Then the current 
\begin{equation}
P^\alpha\equiv \frac{1}{|\sigma|^6}T^{\alpha}{}_{\beta\gamma\delta}\xi^\beta\xi^\gamma\xi^\delta\;, 
\label{eq:ms-conserved-current}
\end{equation}
is conserved in $\mathcal{U}$
\begin{equation}
\nabla_\alpha P^\alpha=0. 
\label{eq:conservation}
\end{equation}
\label{theo:ms-conserved-current}
\end{theorem}
\proof 
To prove this result we first compute the covariant divergence of the superenergy tensor 
$T_{\alpha\beta\gamma\mu}$. We carry out the computation using spinors and hence we need to work out the following 
expression
\begin{equation}
\nabla_{AA'}T^{AA'}{}_{BB'CC'DD'}=4\bar{\mathcal{S}}^{A'}{}_{B'C'D'}\nabla_{AA'}\mathcal{S}^{A}{}_{BCD}+ 
4\mathcal{S}^{A}{}_{BCD}\nabla_{AA'}\bar{\mathcal{S}}^{A'}{}_{B'C'D'}.
\end{equation}
We replace the covariant divergences of the Mars-Simon spinor by the values computed in 
(\ref{eq:covdiv-ms-spinor}) and next use (\ref{eq:ms-superenergy}) on the expression so obtained, getting
\begin{eqnarray}
&&\nabla_{AA'}T^{AA'}{}_{BB'CC'DD'}=
\nonumber\\
&&- \frac{4 \xi^{AA'}}{\sigma}(T_{CB'DC'AD'FA'} \phi_{B}{}^{F} + T_{BB'DC'AD'FA'} \phi_{C}{}^{F} + 
 T_{BB'CC'AD'FA'} \phi_{D}{}^{F})+\mbox{c. c.}
\label{eq:spinor-covdivms-superenergy}
\end{eqnarray}
Next  we compute the covariant divergence of the following quantity
\begin{equation}
Q_{AA'}\equiv T_{AA'BB'CC'DD'} \xi^{BB'} \xi^{CC'} \xi^{DD'}.
\label{eq:spinor-define-q}
\end{equation}
Using eq. (\ref{eq:spinor-covdivms-superenergy}), eq. (\ref{eq:killing-2-form-spinor}), the definition of 
the Ernst potential (\ref{eq:spinor-ernst-potential}) and (\ref{eq:spinor-ernst-1-form})
we get after some algebra 
\begin{eqnarray}
&&\nabla^{AA'}Q_{AA'}=
(T_{AB'BA'CC'DD'} + T_{AB'BC'CA'DD'} + T_{AB'BC'CD'DA'})
\frac{\bar{\sigma}^{A'A} \sigma+\sigma^{AA'}\bar{\sigma}}{\sigma \bar{\sigma}}\times\nonumber\\
&&\xi^{BB'} \xi^{CC'} \xi^{DD'}=3 T_{AA'BB'CC'DD'} u^{AA'}\xi^{BB'} \xi^{CC'} \xi^{DD'}=3 Q_{AA'}u^{AA'}.
\label{eq:spinor-covdivq}
\end{eqnarray}
where in the last step we used the symmetry properties of the spinor $T_{AA'BB'CC'DD'}$ (see \ref{eq:ms-superenergy})
and we set the definition
\begin{equation}
 u_{AA'}\equiv \frac{\bar{\sigma}_{A'A} \sigma+\sigma_{AA'}\bar{\sigma}}{\sigma \bar{\sigma}}=
 \frac{\nabla_\mu (|\sigma|^2)}{|\sigma|^2}.
\label{eq:spinor-u}
\end{equation}
Now, eq. (\ref{eq:spinor-covdivq}) has the tensor representation
\begin{equation}
\nabla_\mu(T^{\mu}{}_{\beta\gamma\delta}\xi^\beta\xi^\gamma\xi^\delta)=
3\frac{\nabla_\mu (|\sigma|^2)}{|\sigma|^2}T^{\mu}{}_{\beta\gamma\delta}\xi^\beta\xi^\gamma\xi^\delta.
\end{equation}
Using this result and the definition of the current $P_\mu$ given by (\ref{eq:ms-conserved-current}) the result follows
after a direct computation.
\qed
\begin{remark}\em
We note that the current $P_\mu$ is defined in terms of the Ernst potential $\sigma$ which in turn 
is defined up to a complex constant (see \ref{eq:define-ernst-potential}) and this means that the current $P_\mu$ indeed represents
a family of currents depending on a complex parameter. Note, however, that the current $P_\mu$ itself is real as is obvious
from (\ref{eq:ms-conserved-current}). Note also that for any point $p\in\mathcal{M}$ we can always choose the complex constant
in such a way that there exists a neighbourhood of $p$ where the Ernst potential does not vanish. Therefore the 
current $P^\mu$ can be always defined locally for any vacuum solution. 
\end{remark}

If a vacuum space-time has a Killing vector then it is very well-known that the Bel-Robinson current has properties similar
to those of the current of Theorem \ref{theo:ms-conserved-current}. The Bel-Robinson current is defined by
\begin{equation}
 (P^{BR})^\mu\equiv B^\mu{}_{\nu\rho\alpha}\xi^\nu\xi^\rho\xi^\alpha\;,
\end{equation}
where $B_{\mu\nu\alpha\beta}$ is the Bel-Robinson tensor (see \cite{SUPERENERGY} for full details). 
The conserved current $P^\mu$ defined in (\ref{eq:ms-conserved-current}) is totally independent from the standard 
Bel-Robinson current. This is more or less apparent from the definition of $P^\mu$
but to make the statement rigorous we compute both currents in a particular example and check explicitly that they differ.
Take the Schwarzschild solution in standard Schwarzschild coordinates $(t,r,\theta,\phi)$
\begin{equation}
 ds^2=\left(1-\frac{2M}{r}\right)dt^2-\frac{dr^2}{(1-\frac{2M}{r})}-r^2(d\theta^2+\sin^2\theta d\phi^2).
\label{eq:schwarzschild}
\end{equation}
If we choose as the Killing vector defining the Mars-Simon tensor
the static Killing vector $\vec{\boldsymbol\xi}=\partial/\partial t$ then a computation reveals
\begin{equation}
\vec{\boldsymbol{P}}=\frac{6 M^2 |k|^2 r (r-2M)}{|k r-2 M|^8}\frac{\partial}{\partial t}\;,\quad
\vec{\boldsymbol{P}}^{BR}=\frac{6 M^2 (r-2M)}{r^7}\frac{\partial}{\partial t}.
\label{eq:schw-current}
\end{equation}
We see from these equations that the currents cannot agree for any value of the complex constant $k$.
\section{Applications}
\label{sec:applications}
The formulation of the non-linear stability problem of the Kerr solution requires the introduction of 
the notion of {\em closeness} of a vacuum initial data set to Kerr initial data. A recent approach towards 
this notion is the concept of {\em non-Kerrness} \cite{KERR-INVARIANT-TJ,KERR-INVARIANT-PRL, 
KERR-INVARIANT-PRS, NON-KERRNESS-COMPACT} which consists in the definition of a non-negative scalar 
quantity from a vacuum initial data set which vanishes if and only if the data are Kerr initial data. In this 
sense the non-Kerrness  enables us to formulate rigorously the non-linear stability problem of the Kerr solution.
However, in order to be able to use the non-Kerrness successfully in any proof of a non-linear stability result,
one needs to have some sort of control on the Cauchy evolution of the non-kerness and this something which so 
far is lacking for any of the notions of non-Kerrness defined so far. We show that this can be remedied 
using the result of Theorem \ref{theo:ms-conserved-current} for a certain
notion of non-Kerness to be introduced next. We start by recalling a local characterisation of the Kerr solution put forward in
\cite{MARS-KERR-UNIQUENESS} and ammended \\ by \cite{MARS-SEN-NMS} 
\footnote{We thank Prof. J. M. M. Senovilla for drawing our attention to \cite{MARS-SEN-NMS}.}.
\begin{theorem}
Let $(\mathcal{M}, g_{\mu\nu})$ be a smooth non-trivial vacuum solution having a Killing vector $\vec{\boldsymbol\xi}$
and assume that it fulfills the following conditions 
\begin{itemize}
\item $(\mathcal{F}_{\mu\nu}\mathcal{F}^{\mu\nu})\neq 0$.
\item There is a choice of the Ernst potential $\sigma$ for which 
\begin{equation}
 \mathcal{S}_{\mu\nu\rho\lambda}=0\;,\quad 
 \mathcal{F}_{\mu\nu}\mathcal{F}^{\mu\nu}+\frac{\sigma^4}{4M^2}=0\;,\quad
M\in\mathbb{R}\setminus\{0\}\;,\quad \mbox{Re}(\sigma)-\lambda >0\;,
\end{equation}
\item There is at least a point $q$ such that the Killing vector $\vec{\boldsymbol\xi}|_q$ 
does not lie in the 2-plane orthogonal to the 2-plane spanned by the two independent null eigenvectors of 
$\mathcal{F}_{\mu\nu}|_q$.
\end{itemize}
Under the previous assumptions the space-time is locally isometric to the Kerr solution.
\label{theo:mars-result}
\end{theorem}
\begin{remark}\em
The result put forward by \cite{MARS-KERR-UNIQUENESS} is in fact more general than
Theorem \ref{theo:mars-result} but we adopt the later particularisation because it is better suited to our
purposes.
\end{remark}

If the space-time is stationary then $\xi^\mu$ is by definition time-like and hence $\xi_\mu\xi^\mu>0$. The mathematical
properties of the superenergy tensor $T_{\mu\nu\alpha\beta}$ (see \cite{SUPERENERGY}) imply that the current $P^\mu$ 
defined by (\ref{eq:ms-conserved-current}) is 
time-like and future-directed if $\xi^\mu$ is future directed too. Therefore, for any embedded 
space-like or null hypersurface $\Sigma\subset\mathcal{M}$ one has 
\begin{equation}
\mathcal{Q}(\Sigma)\equiv\int_{\Sigma} P^\mu n_\mu d\Sigma \geq 0\;, 
\label{eq:global-charge}
\end{equation}
where $n_\mu$ is the (unit) causal future-directed vector orthogonal to $\Sigma$ and $d\Sigma$ is the positive measure on 
$\Sigma$ induced 
by the 3-form $n^\mu\eta_{\mu\alpha\beta\gamma}$\footnote{At a point where $\Sigma$ is null one chooses 
a normal $n^\mu$ in such a way that $n^\mu\eta_{\mu\alpha\beta\gamma}$ is not {\em degenerate}.}. 
Moreover the integral vanishes if and only if the integrand does 
so which can only happen if $P^\mu$ is zero on $\Sigma$ given that $P^\mu$ is time-like and $n_\mu$ causal. 
Using again the mathematical properties of the superenergy tensor \cite{SUPERENERGY} one 
concludes that both $T_{\mu\nu\alpha\beta}$ and $\mathcal{S}_{\mu\nu\alpha\beta}$ vanish on $\Sigma$. 
If $\Sigma$ is space-like then the causal propagation of the Mars-Simon tensor \cite{GAR-VAL-KERR}
enables us to conclude that the Mars-Simon tensor vanish in the Cauchy development of the hypersurface $\Sigma$.
Next assume that $\mathcal{F}_{\mu\nu}\mathcal{F}^{\mu\nu}\neq 0$ and choose the Ernst potential in such a way that on $\Sigma$
\begin{equation}
\left.\left(\mathcal{F}_{\mu\nu}\mathcal{F}^{\mu\nu}+\frac{\sigma^4}{4M^2}\right)\right|_{\Sigma}=0\;,\quad
M\in\mathbb{R}\setminus\{0\}\;,\quad \left(\mbox{Re}(\sigma)-\lambda\right)|_{\Sigma} >0.
\label{eq:ernst-choice}
\end{equation}
The condition $\mathcal{F}_{\mu\nu}\mathcal{F}^{\mu\nu}\neq 0$ implies that $\sigma|_{\Sigma}\neq 0$.
Define the quantity
\begin{equation}
\Xi=\mathcal{F}_{\mu\nu}\mathcal{F}^{\mu\nu}+\frac{\sigma^4}{4M^2}.
\label{eq:define-xi}
\end{equation}
This quantity fulfills the space-time propagation equation (see eq. (4.13) of \cite{GAR-VAL-KERR}) 
\begin{equation}
\nabla_\mu\Xi=-2\mathcal{F}^{\alpha\beta}\mathcal{S}_{\mu\rho\alpha\beta}\xi^\rho+\frac{4\Xi\sigma_\mu}{\sigma}.
\end{equation}
Hence if $\mathcal{S}_{\mu\rho\alpha\beta}=0$ we can integrate the previous equation, getting
\begin{equation}
 \Xi=A \sigma^4\;,\quad A\in\mathbb{C}.
\end{equation}
The assumption (\ref{eq:ernst-choice}) entails $A=0$ and hence $\Xi=0$ on the space-time. The other conditions
of (\ref{eq:ernst-choice}) will also hold in a space-time neighbourhood of $\Sigma$ by continuity. 
The third condition of Theorem \ref{theo:mars-result} is trivially satisfied because the space-time is stationary
and the first condition is fulfilled in a neighbourhood of $\Sigma$ if and only if it holds in $\Sigma$ itself. 
We have thus proven the following result. 
\begin{theorem}
 Let $(\mathcal{M}, g_{\mu\nu})$ be a vacuum stationary solution of the Einstein's field equations
 and assume further that for a given embedded space-like hypersurface $\Sigma$, 
 $(\mathcal{F}_{\mu\nu}\mathcal{F}^{\mu\nu})|_{\Sigma}\neq 0$ and the Ernst potential is chosen in such a way that 
 it fulfills (\ref{eq:ernst-choice}). 
 Use the stationary Killing vector and the Ernst potential to 
 define the vector $P^\alpha$ according to (\ref{eq:ms-conserved-current}). 
 Then the scalar $\mathcal{Q}(\Sigma)$ defined
 by (\ref{eq:global-charge}) is non-negative, it vanishes if and only if $\Sigma$ can be isometrically embedded
 within an open subset of the Kerr solution and it is a conserved charge.
\label{theo:conserved-charge}
 \end{theorem}
\qed

The scalar $\mathcal{Q}(\Sigma)$ can be rendered as an expression involving vacuum initial data if we use the notion of 
{\em Killing initial data set} in the way done in \cite{GAR-VAL-KERR}. 
Therefore we can say that $\mathcal{Q}(\Sigma)$ qualifies as a non-Kerrness scalar defined for a vacuum initial data set.

\section{Conclusions and further perspectives}
\label{sec:conclusions}
We have found a new family of conserved currents for any vacuum solution of the Einstein's equations which is different from those
already known.  The current gives rise to non-negative conserved charges for stationary vacuum solutions which under 
certain circumstances vanish on a space-like hypersurface if and only if the 
hypersurface is a Cauchy hypersurface in a stationary region of the Kerr solution. The existence of this conserved charge 
serves as a complement to the notion of {\em non-Kerrness} because our conserved charge
has positivity properties similar to a non-Kerrness if the conditions described in Theorem \ref{theo:conserved-charge} hold. 
Although these conditions might look somewhat restrictive, we have a control on the Cauchy evolution 
which is not available for the current notions of non-Kerrness.

An interesting open question is the application
of our conserved charge to the problem of the non-linear stability of the Kerr solution. 
The formulation of this problem is, roughly speaking, the proof of the global existence of vacuum data of 
the Einstein's equations which are close to {\em Kerr data} in some sense. The definition of closeness to Kerr 
data can be formulated using the notion of non-Kerrness but to establish the global existence it is necessary 
to control in some way the evolution of the scalar quantity used in the definition of the non-Kerrness. This control is achieved
in our case thanks to the charge conservation.

Another interesting open problem is whether this result can be 
extended in some way for hypersurfaces with a mixed causal character (space-like and null).
This would provide a generalisation of the different notions of {\em non-Kerrness} available in the literature
which are usually formulated on space-like hypersurfaces.
\newpage
\section*{Acknowledgements}
We thank Prof. J. M. M. Senovilla for reading the manuscript and comments. 
Supported by the project FIS2014-57956-P of Spanish ``Ministerio de Econom\'{\i}a y Competitividad'' and 
PTDC/MAT-ANA/1275/2014 of Portuguese ``Funda\c{c}\~{a}o para a Ci\^{e}ncia e a Tecnologia''.

\providecommand{\bysame}{\leavevmode\hbox to3em{\hrulefill}\thinspace}
\providecommand{\MR}{\relax\ifhmode\unskip\space\fi MR }
\providecommand{\MRhref}[2]{%
  \href{http://www.ams.org/mathscinet-getitem?mr=#1}{#2}
}
\providecommand{\href}[2]{#2}

\end{document}